 \documentclass[
reprint,
superscriptaddress,
showpacs,
preprintnumbers,
nofootinbib,
amsmath,
amssymb, 
aps,
prd,
usenames,
dvipsnames,
floatfix
]{revtex4-2}

\pdfoutput=1

\usepackage{mathrsfs}
\usepackage{cancel}
\usepackage{bbold}
\usepackage{makecell}
\usepackage{braket}
\usepackage{bm}
\usepackage{physics}
\usepackage{multirow}
\usepackage{xspace}
\usepackage{newunicodechar}
\newunicodechar{−}{-}
\usepackage{rotating}
\usepackage{ragged2e}
\usepackage{dcolumn}
\usepackage[caption=false]{subfig}  
\usepackage{booktabs}
\usepackage{isotope}

\usepackage{ifpdf}
\ifpdf        
  \usepackage{graphicx, hyperref, xcolor}     
\else     
  \usepackage[dvipdfmx]{graphicx, hyperref, xcolor}     
\fi
 \usepackage[capitalize]{cleveref}

\usepackage{fontawesome} 
\definecolor{blue-violet}{rgb}{0.33, 0.17, 0.89}

\definecolor{rossoferrari}{HTML}{D9073D}
\definecolor{mediumblue}{HTML}{0000CD}
\definecolor{forestgreen}{HTML}{228B22}
\definecolor{desy_blue}{HTML}{009EE2}
\definecolor{desy_orange}{HTML}{FD8800}
\hypersetup{
setpagesize=false,
bookmarksnumbered=true,%
bookmarksopen=true,%
colorlinks=true,%
linkcolor=rossoferrari,
urlcolor=mediumblue,
citecolor=mediumblue,
linktocpage=false
}

\newcommand{\xseventeen}{$X_{17}$\xspace}

\makeatletter
\@addtoreset{equation}{section}

\makeatother

\usepackage{tikz}
\usepackage{tikz-feynman}
\tikzfeynmanset{compat=1.0.0}

\newcommand{\beq}{\begin{equation}}
\newcommand{\eeq}{\end{equation}}
\newcommand{\be}{\begin{equation}}
\newcommand{\ee}{\end{equation}}

\begin{document}

\title{Kaon decay constraints on vector bosons coupled to non-conserved currents}

\author{Matheus Hostert}
\email{matheus-hostert@uiowa.edu}
\affiliation{Department of Physics and Astronomy, University of Iowa, Iowa City, IA 52242, USA}

\author{Maxim Pospelov}
\email{pospelov@umn.edu}
\affiliation{School of Physics and Astronomy, University of Minnesota, Minneapolis, MN 55455, USA}
\affiliation{William I. Fine Theoretical Physics Institute, School of Physics and Astronomy, University of Minnesota, Minneapolis, MN 55455, USA}

\author{Adrian Thompson}
\email{a.thompson@northwestern.edu}
 \affiliation{Department of Physics \& Astronomy, Northwestern University, 2145 Sheridan Road, Evanston, IL 60208, USA}

\date{\today}

\begin{abstract}
We study rare three- and four-body kaon decays as a probe of light vector and axial-vector bosons coupled to non-conserved currents.
We find that searches for $K_L \to \pi^0 \pi^0 (X\to e^+e^-)$ decays constrain the couplings of light $X$ bosons to light quarks to be as small as $\mathcal{O}(10^{-5})$.
The charged-pion modes $K^+ \to \pi^+ \pi^0 (X \to e^+e^-)$ and $K_L \to \pi^+ \pi^- (X \to e^+e^-)$ provide weaker limits, but constrain complementary combinations of couplings to the $u$, $d$, and $s$ quarks at the level of $\mathcal{O}(10^{-4})$.
Finally, we also find that double emission of $X$ in $K \to \pi XX$ decays can provide yet additional constraints on the parameter space of light $X$ bosons due to a double $(m_K/m_X)^2$ enhancement to the rate.
For a 17 MeV boson, these limits add to the known tension between spin-1 bosons coupled to vector and axial-vector currents interpretations of the results of the ATOMKI experiment with meson decay data.
Finally, we also comment on negative pion capture on hydrogen and deuterium as a source of light particles and discuss the prospects for testing the 17 MeV boson hypothesis. 
\end{abstract}

\maketitle

{
\hypersetup{hidelinks} 
\tableofcontents
}

\section{Introduction}
\label{sec:intro}

The existence of new force carriers has been a subject of intense theoretical and experimental interest in recent years~\cite{Agrawal:2021dbo,Antel:2023hkf}.
This is particularly true for MeV- and GeV-scale bosons, which can be searched for under experimental lampposts like high-intensity beam dumps and in the rare decays of mesons~\cite{Goudzovski:2022vbt,Aebischer:2025mwl}.
In the case of vector particles, the most well-studied example is the dark photon~\cite{Holdom:1985ag}, which couples to the electromagnetic current with a strength suppressed by the small kinetic mixing parameter $\varepsilon$.
However, while a compelling and minimal model, the dark photon is not the only possibility, and many other scenarios have been discussed. 
If such a boson is associated with a local gauge symmetry, then gauge invariance requires that it couples to a conserved and anomaly-free current.
As far as its couplings to Standard Model (SM) fields are concerned, this implies that the new vector boson $X_\mu$ should couple to linear combinations of the electromagnetic current $J_\mu^{\rm EM}$ and the baryonic and leptonic currents $J_\mu^{B_i}$ and $J_\mu^{L_i}$ with $i$ a SM generation index.

Vectors coupled to other more exotic and non-conserved currents, including fully axial-vector couplings, can still be considered as low-energy effective theories~\cite{Preskill:1990fr} on which we can place quite stringent experimental limits.
As discussed in ~\cite{Dror:2017ehi,Dror:2017nsg}, this is because the effective theory of a vector boson coupled to non-conserved currents features enhanced production of the longitudinal mode of $X_\mu$.
This is also the case for SM extensions that include anomalons: heavy fermions that cancel anomalies at or above the electroweak scale.
These generate effective Wess-Zumino terms at low energies that lead to a similar behavior~\cite{DHoker:1984izu,DHoker:1984mif}. 

If the new vector boson couples to a non-conserved current, the longitudinal mode of $X_\mu$ can be produced in meson decays with an enhancement factor of $(E/m_X)^2$, where $E$ is the typical energy of $X_\mu$ in the process and $m_X$ is the mass of the new vector boson~\cite{Fayet:2007ua,Pospelov:2008zw}.
This is easy to see with the Goldstone-boson equivalence $X_\mu \to \frac{\partial_\mu \phi}{m_X}$ relation, as the Goldstone mode $\phi$ couples as
\begin{equation}
        \mathcal{L} \supset g_X X_\mu J^\mu \longrightarrow -\frac{g_X}{m_X}\phi\,  \partial_\mu J^\mu,
\end{equation}
which in the absence of current conservation, $\partial_\mu J^\mu \neq 0$, leads to an $E/m_X$-enhanced coupling for light bosons.

One phenomenological context for such exotic vectors is the now long-standing anomalies in rare nuclear transitions at the ATOMKI experiment.
Several anomalies have been reported by the collaboration including in nuclear transitions of $\isotope[8]{Be}$~\cite{Krasznahorkay:2018snd,Krasznahorkay:2019lyl}, $\isotope[4]{He}$~\cite{Krasznahorkay:2021joi}, and $\isotope[12]{C}$~\cite{Krasznahorkay:2022pxs}.
These constitute a large excess of $e^+e^-$ pairs with large opening angles in rare nuclear de-excitations.
The new-physics explanation is based on the emission of a boson of mass around $17$~MeV, dubbed here $X_{17}$, which subsequently decays to $e^+e^-$. 
Because the boson mass is very close to the transition energy, it is emitted with small velocity, explaining the excess of nearly back-to-back $e^+e^-$ pairs.
Several models have been proposed in the literature to explain these results, including exotic vector bosons~\cite{Feng:2016jff,Feng:2016ysn,Feng:2020mbt}, MeV-scale QCD axions~\cite{Alves:2017avw,Alves:2020xhf,Alves:2024dpa}, and scalar or pseudo-scalar particles~\cite{Barducci:2022lqd}. At the most basic level one could question if any hypothetical $X_{17}$, with arbitrarily adjusted couplings to quarks and leptons, could be consistent with available data. 
One finds that all models are strongly constrained by data, including searches for new particles in meson decays, $e^+e^-$ colliders, beam dumps, and fixed target experiments like NA64~\cite{NA64:2018lsq,NA64:2019auh,NA64:2021aiq}.

Meson decays, in particular, set stringent constraints on all such scenarios since the \xseventeen should couple to quarks and decay relatively fast to $e^+e^-$.
For instance, meson decays such as $\pi^0 \to \gamma (X_{17} \to e^+e^-)$ exclude the possibility that this particle is a dark photon~\cite{NA482:2015wmo}.
Moreover, this constraint combined with limits on charged pion decays $\pi^+ \to e^+\nu_e (X_{17} \to e^+e^-)$ exclude the possibility that \xseventeen has vector coupling to quarks~\cite{Hostert:2023tkg}, and with the calculations of the pure axial-vector boson in \cite{Mommers:2024qzy}, they also exclude the required couplings to explain the $\isotope[12]{C}$ results.
It also places strong limits on the 17 MeV axion~\cite{Alves:2017avw,Alves:2020xhf,Alves:2024dpa}, which is also excluded by $K\to \pi(a \to e^+e^-)(a\to e^+e^-)$ searches at NA62~\cite{Hostert:2020xku,NA62:2023rvm}. 

In this article, we revisit kaon decays as a powerful probe of a vectorial $X$ coupled to flavor-conserving vector and axial-vector couplings of light quarks. 
Several of the channels discussed here were investigated in detail for MeV QCD axions in Refs.~\cite{Alves:2017avw,Alves:2020xhf,Alves:2024dpa}, but had not been studied for vector bosons since they are typically subdominant in scenarios with conserved currents.
We focus on decays of the type $K\to X_{ee}+\dots$, with $X_{ee}$ indicating a new vector that decays to $e^+e^-$.
Of greatest relevance is the branching ratio for $K_L \to \pi^0\pi^0 X_{ee}$ channel which is constrained at the level of $\mathcal{O}(10^{-8})$.
We also comment on doubly-enhanced $X_\mu$ emission in $K\to \pi X_\mu X_\mu$ and on other future opportunities with $\pi^-$ capture.
As we will see, kaon decays constrain several different combinations of strange, up, and down quark couplings, which, combined, leave even less room for \xseventeen particles.
Indeed, given the multitude of meson limits and the required vector and axial-vector couplings required to explain the ATOMKI results, we find it extremely unlikely that the \xseventeen particle is the correct explanation for their results.

This article is divided as follows. In \cref{sec:model} we discuss our model and formalism used to calculate the kaon decays.
The latter are discussed in \cref{sec:kaons} and we present and discuss our main results in the context of \xseventeen in \cref{sec:results}.
Finally, in \cref{sec:pions} we discuss additional future directions that could further explore the \xseventeen hypothesis, and conclude in \cref{sec:conclusions}.

\section{Vector Bosons}
\label{sec:model}
 
Our starting point will be the vector and axial-vector couplings of $X$ to quarks.
Using SU$(3)$ chiral perturbation theory (ChPT), we then derive all the relevant couplings to pions and kaons (see \cref{app:chpt}), from which we calculate the new partial widths of $K^\pm$, $K_L$, $K_S$, and others.
Our interaction Lagrangian for a vector/axial-vector $X$ coupling to fermions $f$ includes only up, down, or strange quark fields, as well as the electron,
\begin{equation}
    \mathcal{L} \supset X_\mu \left(\sum_{f=u,d,s,e} \bar{f} (g^V_f \gamma^\mu + g_f^A \gamma^\mu \gamma^5 ) f\right).
\end{equation}
These then enter into the ChPT Lagrangian through operator matching.

\begin{figure}[t]
    \centering
    \includegraphics[width=0.5\linewidth]{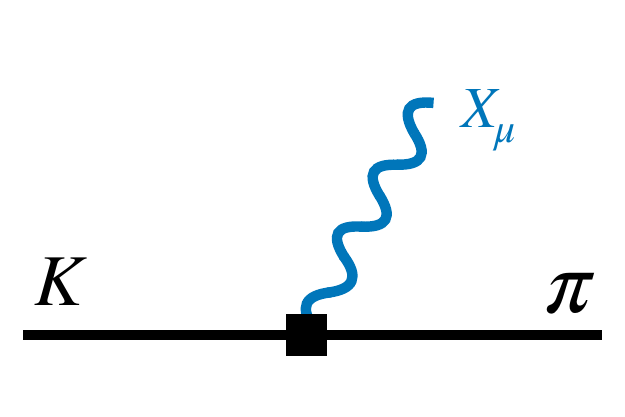}
    \caption{Leading diagram for $K^+\to \pi^+ X_\mu$ and $K_S \to \pi^0 X_\mu$ decays.}
    \label{fig:KtoPiX_diagram}
\end{figure}

The $\Delta S = 1$ strangeness violating interaction~\cite{Cirigliano:2011ny} contains
\begin{equation}
\label{eq:chpt_deltaS1}
\mathcal{L}^{\Delta S = 1}_{G_8 p^2} = F^4 G_8 \langle U_- (D_\mu U)^\dagger  D^\mu U\rangle + \rm h.c.
\end{equation}
where $\langle\,\dots\rangle$ denotes the trace, $G_8$ is the coupling (mass dimension $-2$) for this operator transforming under the octet, $F \simeq 93$ MeV is the pion decay constant, 
$U_- \equiv (\lambda_6 - i \lambda_7) / 2$ is a linear combination of Gell-Mann matrices to pick out $\bar{s} \to \bar{d}$ transitions, and $U = e^{i \phi / F}$ contains the $3\times3$ meson octet $\phi$.
The covariant derivatives take the form
\begin{align}
    D_\mu U &= \partial_\mu U - i X_\mu [G_V, U ] & \text{vector} \nonumber 
\\
    D_\mu U &= \partial_\mu U - i X_\mu \{G_A, U \} & \text{axial-vector}
\end{align}
for $G_{V,A} = \, -\text{diag}({g^{V,A}_u,g^{V,A}_d,g^{V,A}_s})$. 
Expanding \cref{eq:chpt_deltaS1} to fourth order in meson fields reveals the operators necessary to support $K_{L,S} \to \pi X$, $K_L \to \pi^0 \pi^0 X$, $K_L \to \pi^+ \pi^- X$, $K^+ \to \pi^0 \pi^+ X$, and other reactions which we now discuss. 
In this work, we will also express the $u$, $d$, and $s$ couplings in the basis of $T$, $U$, $V$ flavor subgroup couplings, given in \cref{tab:dict}, when they offer more compact amplitude expressions. 
We note in passing that in this approach a model with $G_A=0$ and $G_V$ proportional to the unit matrix in flavor space gives $D_\mu U= \partial_\mu U$. 
However, even in this model the two-loop anomalous diagrams lead to effective coupling of $X_\mu$ to flavor-changing $s$-$d$ current \cite{Dror:2017ehi,Dror:2017nsg}. 
In this paper we neglect these effects that depend on the couplings to heavier quarks. 
In principle, one could include them in our analysis as well, and it would amount to postulating an off-diagonal in flavor space coupling $g_{ds}^{V-A}$ with a much smaller magnitude than $g^{V,A}_{u,d,s}$ input. 

Before proceeding, we recall that the $\Delta S=1$ operators are sourced by weak interactions, so it is often convenient to describe their effect via the small dimensionless ratio
\begin{equation} \label{eq:kappa_8}
    \kappa_8 = G_8 F^2 \simeq - 7.7 \times 10^{-8}.
\end{equation}
In particular, the operator in \cref{eq:chpt_deltaS1} induces mixed kaon-pion and kaon-eta propagators, which we remove by diagonalizing the meson kinetic terms.
In what follows, we work in this new basis which is the physical basis up to small $\kappa_8^2$ corrections.

We concentrate on $G_8$-proportional operators, as their dominance over other isospin structures such as $G_{27}$ is a well-established experimental fact (see \cite{Cirigliano:2011ny} and references therein). For that reason, we are allowed to neglect $G_{27}$, and base our predictions only on the $G_8$-proportional interactions. Moreover, in some channels suppressed by $\Delta T = 1/2 $ selection rule such as $K^+\to\pi^+\pi^0$, the emission of an extra $X$ particle would circumvent the suppression, allowing $G_8$-proportional effects. (A similar phenomenon occurs with radiative modes where $K^+\to \pi^+\pi^0 \gamma$ operators is dominated by the $G_8$ coupling.) We also limit our discussion to the lowest order ChPT calculation. This does not have a solid justification as $m_K/(4\pi F) \simeq 0.4$, and subleading ChPT contributions could be important. On the other hand, there is a significant number of sensitive kaon decay modes and ``accidental" cancellation of leading and subleading orders for all of them is not feasible. 

\section{Kaon decays}
\label{sec:kaons}

\subsection{$K\to \pi X_{ee}$ decays}

For completeness, we start with the two-body decays of kaons to a single dark vector.
As we will see, these usually do not impose strong limits for $m_X < m_{\pi^0}$ due to the large Dalitz backgrounds from $K\to \pi \pi^0_D$.
In addition, for on-shell $X$, the decay must involve only the longitudinal mode of $X$ due to the conservation of angular momentum.
Off-shell decays are allowed, but will be further suppressed by the leptonic couplings of $X$.

The electromagnetic three-body decays of $K_S$ are constrained by NA48~\cite{NA481:2003cfm}
\begin{equation}
    \mathcal{B}(K_S \to \pi^0 e^+e^-, m_{ee} > 165\text{ MeV}) = (3.0 ^{+1.5}_{-1.2})\times 10^{-9}
\end{equation}
An earlier NA48 analysis~\cite{NA48:2001kje} does not impose the stringent $m_{ee}$ cut as in~\cite{NA481:2003cfm} and can be used to set limits on the final state $X\to e^+e^-$.
For $K_L$, the leading limits come from KTEV and give~\cite{KTeV:2000hbv,KTeV:2003sls},
\begin{equation}
    \mathcal{B}(K_L \to \pi^0 e^+e^-) < 2.8 \times 10^{-10} \text{ at 90\% CL},
\end{equation}
where $140 \text{ MeV} < m_{ee} < 362.7$~MeV was imposed to reduce Dalitz backgrounds.
Finally, NA48/2 measured $K^+ \to \pi^+ e^+e^-$ to great precision~\cite{NA482:2009pfe},
\begin{equation}
    \mathcal{B}(K^+\to \pi^+e^+e^-) = (3.11 \pm 0.12) \times 10^{-7},
\end{equation}
in the full kinematic range, in agreement with E949 measurement~\cite{E865:1999ker}. All analyses impose a cut on $m_{ee} > 140$~MeV.

For lighter bosons, much older analyses exist~\cite{Yamazaki:1984vg,Baker:1987gp}.
The analysis in \cite{Baker:1987gp}, for example, concludes that $\mathcal{B}(K^+\to \pi^+ X_{ee}) < 4.5 \times 10^{-7}$ for $\tau(X) < 10^{-13}$~s and $m_X = 1.8$~MeV.
However, Ref.~\cite{Alves:2017avw} raises several caveats about these limits, arguing instead for a more conservative limit of
\begin{equation}\label{eq:KpiX_limit}
    \mathcal{B}(K^+\to \pi^+ X_{ee}) \lesssim 10^{-5}.
\end{equation}
We reiterate the point in \cite{Alves:2017avw} that modern bump hunts in $K\to \pi e^+e^-$ would likely bring tighter constraints.

\begin{figure}[t]
    \centering
    \includegraphics[width=\linewidth]{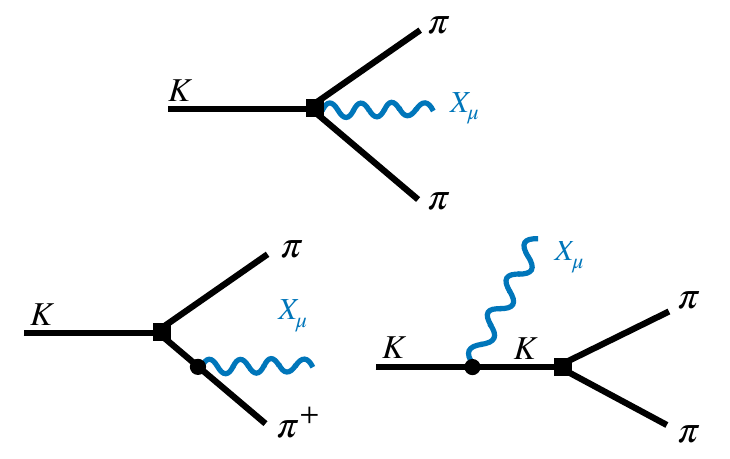}
    \caption{Leading diagrams for $K_L\to \pi^0 \pi^0 X_\mu$, $K^+\to \pi^+ \pi^0 X_\mu$, and $K_L \to \pi^+ \pi^- X_\mu$ decays. Emissions from a single pion line or from a three-pion vertex is only significant for modes with charged pions.}
    \label{fig:KtoPiPiX_diagram}
\end{figure}

The $K\to \pi X$ decays can proceed via $\Delta S = 1$ tree-level transitions in ChPT (c.f. \cref{eq:chpt_deltaS1}) or via loop-level FCNC decays.
As already mentioned, if the new vector couples to top or charm quarks, the latter can dominate the new physics rate for non-conserved currents, see {\em e.g.} Refs.\,\cite{Dror:2017ehi,Dror:2017nsg}. 
Here, we focus instead on the tree-level emission calculated in ChPT, as shown in \cref{fig:KtoPiX_diagram}.
The conservative limit in \cref{eq:KpiX_limit} will constrain a unique combination of couplings of \xseventeen, but at a smaller order of magnitude than the three-body decays we discuss next.

With the Lagrangian in \cref{eq:XpK}, we find
\begin{align}
    \Gamma_{K_S \to \pi^0 X} &= \frac{\kappa_8^2 m_{K}}{\pi} \frac{|\vec{p}_X|^2}{m_X^2} (g^X_{K\pi^0})^2 \frac{|\vec{p}_X|}{m_K},
    \\
    \Gamma_{K^+ \to \pi^+ X} &= 
  \frac{\kappa_8^2 m_{K}}{4\pi} \frac{|\vec{p}_X|^2}{m_X^2} (g^X_{K\pi^+})^2 \frac{|\vec{p}_X|}{m_K}
\end{align}
where we define the coupling combinations, using Table~\ref{tab:dict} to translate between the $u$, $d$, $s$ and the flavor subgroup $T$, $U$, $V$ basis, as
\begin{align}
    \label{eq:g_kpineutral}
    g_{K\pi^0}^X &\equiv g_{U_0}^A + g_{U_3}^R + g_{U_3}^V\left(\frac{r}{1-r}\right)  \nonumber
    \\
    &=\frac{1}{2}\bigg(3 g_d^A - g_d^V + g_s^A + g_s^V \bigg) + \frac{g_d^V - g_s^V}{1-r} \\
    \label{eq:g_kpi}
    g_{K\pi^+}^X &\equiv g^A_{T_0} + g^A_{V_0} + (g^V_{V_3} - g^V_{T_3}) \left(\frac{1 + r}{1 - r}\right) \nonumber \\
    &=2 g_u^A + g_d^A + g_s^A + (g_d^V - g_s^V)\frac{1+r}{1-r} \, ,
\end{align}
where we defined the mass ratio
\begin{equation}
    r = \frac{m_\pi^2}{m_K^2},
\end{equation}
and the chiral couplings,
\begin{equation}
    g_i^L = \frac{g_i^V-g_i^A}{2}, 
    \quad 
    g_i^R = \frac{g_i^V+g_i^A}{2}.
\end{equation}
The $K_L$ decay is CP-violating and, hence, neglected here.
Both $K_S$ and $K^+$ decays receive contributions from $K-\pi-X$ vertex as well as from $K-\pi$ mixing alongside $X$ emission from a vector current.
We neglect diagrams with eta mesons, which can contribute to the $K_S$ decay through mixing.
Note that the $K^+\to \pi^+ X$ decay vanishes in the limit of current conservation, $g_{V_3}^V = g_{T_3}^V$ (i.e., $g_s^V = g_d^V$).
Conserved currents, such as the case of dark photon $X$ coupled to electromagnetic current \cite{Pospelov:2008zw} would generally require a loop-level interaction similar to $K\pi\gamma^*$ effective vertex \cite{DAmbrosio:1998gur}.

To summarize this subsection, in order to employ the tightest bounds on $X$ using $K\to \pi X_{ee}$ modes, one would have to restrict $X$ to mass range above the pion threshold,
in order to avoid excessive $\pi^0$ Dalitz background. Of course this is not ideal for $X_{17}$ searches, and one can circumvent these limitations by utilizing higher multiplicity modes as discussed below.

\subsection{$K_L \to \pi^0\pi^0 X_{ee}$ decays}

In this section we compute the emission of the new bosons in $K \to \pi^0 \pi^0 X_{ee}$ decays. 
This channel is particularly sensitive to new physics because the all-neutral mode $K_L \to \pi^0 \pi^0 e^+e^-$ is suppressed with respect to its charged counterparts we consider below. 
In the SM, $K_L \to 2\pi$ decays require CP violation, and, in addition, the electromagnetic $K_L\to 2\pi^0 \gamma$ channel lacks the internal bremsstrahlung diagrams available in the charged modes.
The ChPT estimate of Ref.~\cite{Heiliger:1993qi} accounts for emission of virtual photons and an E1 transition to give
\begin{equation}
    \mathcal{B}(K_L\to\pi^0\pi^0 e^+e^-) \simeq 2 \times 10^{-10}.
\end{equation}
The KTeV experiment searched for this channel~\cite{KTeV:2002tpo}.
Having observed a single event, they set an upper limit on the branching ratio of
\begin{equation}
    \mathcal{B}(K_L \to \pi^0 \pi^0 e^+e^-) < 6.6 \times 10^{-9} \text{ at 90\% CL},
\end{equation}
with no direct cuts on the invariant mass of the $e^+e^-$ pair.
As we will see, this channel will provide a powerful limit on light vectors and axial-vectors coupled to non-conserved currents.

We find that the direct $X$ diagram (the contact term in \cref{fig:KtoPiPiX_diagram}) is proportional to
\begin{align}\label{eq:g_kpipiX_neutral}
    g_{K\pi^0\pi^0}^X &\equiv g_{U_3}^R+g_{U_0}^A - g_{U_3}^A \frac{r}{1 - r} \nonumber \\
    &=\frac{1}{2} \bigg( g_d^V + 3 g_d^A - g_s^V + g_s^A\bigg) + r\frac{g_s^A - g_d^A}{1-r} \, .
\end{align}
At $\mathcal{O}(p^2)$, the only other possibility is the emission of $X$ from the kaon line, turning it into $K_S$, which then enables the CP-conserving decay to two pions.

For a 17 MeV boson, including both the contact diagram and the bremsstrahlung emission from the kaon line, we find
\begin{align}
    \mathcal{B}(K_L &\to \pi^0 \pi^0 X_{17}) \simeq  42.3 \times \left(\frac{17 \text{ MeV}}{m_X}\right)^2 \nonumber \\
    &\times \bigg(0.22 (g_{K\pi^0\pi^0}^X)^2 - 0.75 g_{K\pi^0\pi^0}^X g_{U_3}^V + (g_{U_3}^V)^2\bigg).
\end{align}

\subsection{$K^+ \to \pi^+\pi^0 X_{ee}$ decays}
Now moving to charged kaon decays, we must resort to a bump hunt to search for $X_{ee}$.
The decay of the charged kaon to pions and electron-positron pairs was measured by NA48/2~\cite{NA482:2018gxf}, with branching ratio
\begin{align}
    \mathcal{B}(K^+ \to \pi^+ \pi^0 e^+e^-) &= (4.24 \pm 0.14) \times 10^{-6}  \, .
\end{align}
In order to estimate the exclusionary power of the NA48/2 data over a $K^+ \to \pi^+ \pi^0 X_{ee}$ signature, we construct a binned $\chi^2$ over the data and backgrounds in the invariant mass of the electron-positron system, $m_{ee}$.

The amplitude for $K^+ \to \pi^+ \pi^0 X$ is comprised of three interfering matrix elements given by the topologies shown in \cref{fig:KtoPiPiX_diagram}. 
However, the diagram with $X$ emission off the kaon leg is suppressed by the charged and neutral pion mass difference, $(m_{\pi^+}^2 - m_{\pi^0}^2)$, vanishing in the SU$(2)$ limit and only making its appearance suppressed by the 27-plet coupling $G_{27}$ in the ChPT Lagrangian\, \cite{Donoghue1992Dynamics}. 
Therefore, we approximate the amplitude for this channel by only taking the interference of the contact and pion emission diagrams.

From the contact vertex given in \cref{eq:XppK}, the $\pi \pi X$ vertex in Eq.~\ref{eq:PiPiX} (proportional to $g_{T_3}^V$), and the $K \pi \pi$ operator in the SM, the decay width for a 17 MeV boson is found to be
\begin{align}
\label{eq:KPlusPiPlusPi0_couplings}
\mathcal{B}(K^+ &\to \pi^+ \pi^0 X) \simeq \bigg(\frac{17\, \,  \rm{MeV}}{m_X} \bigg)^2 \\\nonumber
&\times \bigg( a \big(g_{T_3}^V - g_{T_3}^A/3\big)^2 + b (g_{K\pi^+\pi^0}^X)^2   \\\nonumber
& + c (g_{T_3}^V)^2 + d g_{T_3}^V \big(g_{T_3}^V - g_{T_3}^A/3\big)
 + e g_{T_3}^V g_{K\pi^+\pi^0}^X \bigg)\, ,
\end{align}
where the numerical constants $(a,b,c,d,e) \simeq (1.1, 0.13, 4.9, -4.4, 0.51)$ and the term proportional to $b$ is the contact coupling in \cref{eq:XppK},
\begin{align}\label{eq:Kplus_piplus_pi0_X}
    g_{K\pi^+\pi^0}^X &\equiv g_{V_3}^V+ g_{T_0}^A + g_{U_0}^A + 2\frac{g_{T_3}^A - g_{V_3}^A r}{1-r}  \nonumber \\
    &= g_s^A - g_s^V + g_u^A + g_u^V+ 2\frac{g_u^A - (g_d^A - g_s^A + g_u^A)r}{1 - r}.
\end{align}
We then determine the signal yield for $1.7 \times 10^{11}$ observed $K^+$ decays. Additionally, we apply a Gaussian smearing of the signal $m_{ee}$ distribution with resolution adopted from ref.~\cite{NA482:2015wmo}. The acceptances as a function of $m_{ee}$ are reported in ref.~\cite{NA482:2018gxf} over three separate contributions to the amplitude structure of the $K^+ \to \pi^+ \pi^0 e^+ e^-$ signal, the internal bremsstrahlung ``IB'' and magnetic ``M'' contributions, in addition to an interference term contribution which does not apply here. While the $K^+ \to \pi^+ \pi^0 X$ axial-vector emission is most physically similar to the magnetic terms, we consider exclusions under the acceptances for each test hypothesis. By constructing a binned $\chi^2$ test statistic over the $m_{ee}$ data, also including $m_{ee}$-independent systematic and external errors added in quadrature to the bin-by-bin statistical uncertainties, we extract the resulting constraints on the branching ratio for $K^+ \to \pi^+ \pi^0 X_{ee}$ at $m_X = 17$ MeV, which are
\begin{align}
    \mathcal{B}(K^+ &\to \pi^+ \pi^0 X_{ee}^{17}) \, < 1.89 \times 10^{-8} &\text{(IB)}\,\nonumber \\
    \mathcal{B}(K^+ &\to \pi^+ \pi^0 X_{ee}^{17}) \, < 1.15 \times 10^{-8}  &\text{(M)} \, ,
\end{align}
which can be interpreted as bounds on the effective coupling combination in \cref{eq:KPlusPiPlusPi0_couplings} at the level of $\mathcal{O}(10^{-4})$ in the absolute values of the individual couplings.

\subsection{$K_{L} \to \pi^+\pi^- X_{ee}$ decays}

We find that this channel is sensitive to the coupling combination,
\begin{align}\label{eq:g_kpipiX}
    &g_{K\pi^+\pi^-}^X \equiv (2 g_{T_3}^R - g_{U_3}^R) - (g_{U_0}^A + g_{T_3}^A) - \frac{g_{T_3}^A - r  g_{V_3}^A}{1 - r} \nonumber \\
    &=\frac{1}{2} \left(g_s^V + 2 g_u^V - 3(g_d^A + g_d^V) - g_s^A\right) 
    \\\nonumber&\quad\quad\quad\quad\quad\quad\quad
    - \frac{g_u^A -g_d^A + (g_s^A-g_u^A) r}{1 - r}
\end{align}
from the contact operator in \cref{eq:operator_kpipiX}. In addition, this channel is sensitive to $g_{T_3}^V$ and $g_{U_3}^V$ from the diagrams in \cref{fig:KtoPiPiX_diagram} with emission of $X$ from the pion and kaon legs, respectively. 
This time the emission from the kaon leg is unsuppressed, unlike the $K^+ \to \pi^+ \pi^0 X$ channel. The branching ratio close to $m_X = 17$~MeV is found to be
\begin{align}
\label{eq:KLPiPlusMinus_couplings}
    \mathcal{B}(K_L &\to \pi^+ \pi^- X) \simeq 227 \times \bigg(\frac{17\rm \, MeV}{m_X}\bigg)^2 \nonumber \\
    \times & \bigg( \tilde{a} (g_{K\pi^+\pi^-}^X)^2 + \tilde{b} (g_{T_3}^V)^2 + \tilde{c} (g_{U_3}^V)^2 \nonumber \\
    &+ \tilde{d} g_{K\pi^+\pi^-}^X g_{T_3}^V + \tilde{e} g_{K\pi^+\pi^-}^X g_{U_3}^V + \tilde{f} g_{T_3}^V g_{U_3}^V  \bigg) \, ,
\end{align}
where $(\tilde{a},\tilde{b},\tilde{c},\tilde{d},\tilde{e},\tilde{f}) \simeq (0.08, 0.74, 0.37, 0.43, 0.28, 1.0)$.

The charged-pion decays of $K_S$ and $K_L$ were measured by NA48~\cite{NA48:2003pwz},
\begin{align}
        \mathcal{B}(K_L \to \pi^+ \pi^- e^+e^-) &= (3.08 \pm 0.20) \times 10^{-7},
        \\
        \mathcal{B}(K_S \to \pi^+ \pi^- e^+e^-) &= (4.71 \pm 0.32) \times 10^{-5}.
\end{align}
In principle, these can also be used to set weaker limits by inspecting the $m_{ee}$ distribution of events.

\begin{figure}[t]
    \centering
    \includegraphics[width=\linewidth]{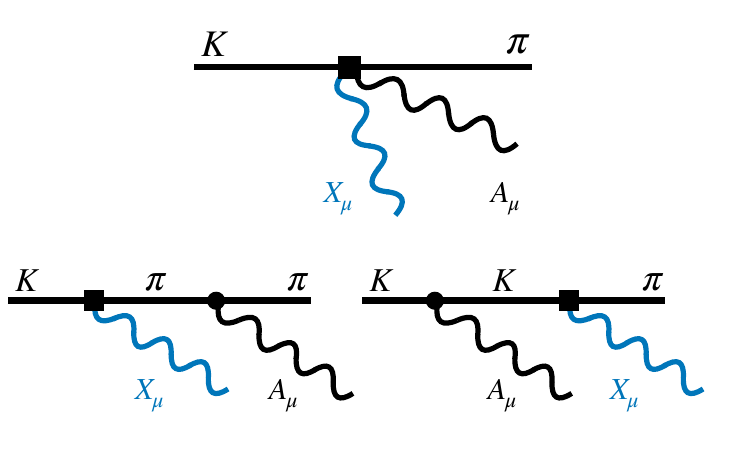}
    \caption{Leading diagrams for $K^+\to \pi^+ \gamma  X_\mu$. 
    For $K \to \pi X_\mu X_\mu$ there are two additional possibilities by swapping the vector and $\Delta S  =1$ vertices of the bottom diagrams.
    }
    \label{fig:KtoPiXgamma_diagram}
\end{figure}

\renewcommand{\arraystretch}{1.5}
\begin{table*}[]
    \centering
    \begin{tabular}{|c|c|c|c|c|}
    \hline
        Channel  & Pure $V$ Couplings & Pure $A$ Couplings & \makecell{SU(3) Limit \\ $g^{V,A}_u = g^{V,A}_d = g^{V,A}_s = g^{V,A}$} & BR 90\% U.L. @17 MeV \\
        \hline
         $K_L \to \pi^0 \pi^0 X_{ee}$  & $g_{U_3}^V$ & $g_{U_0}^A + g_{U_3}^A \frac{1 - 2r}{1-r}$ & $g_{U_0}^A \to g_{\rm eff} = 2 g^A$ & $6.6 \times 10^{-9}$ \\
         \hline
         $K_L \to \pi^+ \pi^- X_{ee}$ & $(2 g_{T_3}^V - g_{U_3}^V) , g_{U_3}^V , g_{T_3}^V$ & $g_{U_0}^A + \mathcal{O}(r)$ & $g_{U_0}^A \to g_{\rm eff} = 2 g^A$  & $2.18 \times 10^{-8}$\\
         \hline
         \makecell{$K^+ \to \pi^+ X_{ee}$, \\ $K^+ \to \pi^+ \gamma X_{ee}$}  & 
         $g_{U_3}^V$ 
         & $g_{T_0}^A + g_{V_0}^A$ & $g_{T_0}^A + g_{V_0}^A \to g_{\rm eff} = 4 g^A$ & \makecell{$10^{-5}$, \\ $1.5 \times 10^{-9}$}\\
         \hline
         $K^+ \to \pi^+ \pi^0 X_{ee}$  & $g_{T_3}^V,\,\, g_{V_3}^V$ & \makecell{$-g_{T_3}^A / 3$, \\ $g_{T_0}^A + g_{U_0}^A + 2 \frac{g_{T_3}^A - g_{V_3}^A r}{1-r}$} & $g_{T_0}^A + g_{V_0}^A \to g_{\rm eff} =  4 g^A$ & $[1.15,1.89] \times 10^{-8}$\\
         \hline
         $K^+ \to \pi^+ X_{ee} X_{ee}$ & \makecell{$(g_{V_3}^V - g_{T_3}^V)^2 \frac{1 + r}{1 - r}$, \\ $-4 \frac{(g_{T_3}^V)^2 + (g_{V_3}^V)^2 r}{1-r} $} & $-(g_{U_0}^A)^2 + 4 \frac{(g_{T_3}^A)^2 + (g_{V_3}^A)^2 r}{1 - r}$ & \makecell{$g_{T_0}^V g_{V_0}^A - (g_{U_0}^A)^2$ \\ $ \to g_{\rm eff}^2 = 4 g^A (g^V - g^A)$} & $3.14 \times 10^{-9}$ \\
         \hline
    \end{tabular}
    \caption{Summary of exclusions and flavor combination for some of the kaon decay channels considered here. A ``$,$'' indicates interference between couplings through different amplitudes, and $r \equiv (m_\pi / m_K)^2 \simeq 0.07$. In the second and third columns, we report the dependence if purely vector or axial-vector couplings are turned on. In the case that both $V$ and $A$ couplings are present, more interference terms appear. Flavor universality takes all triplet couplings to vanish, leaving the resulting coupling combinations in the fourth column. In the last column we report the upper limit on the branching ratio at $m_{ee} = 17$ MeV.}
    \label{tab:constraints}
\end{table*}

We find a similar limit by examining $K_L \to \pi^+ \pi^- e^+ e^-$ decay data at NA48/2~\cite{NA48:2003pwz}. The measured branching ratio is found by normalizing to the observed rate of $K_L \to \pi^+ \pi^- \pi^0_D$ decays, where ``$\pi^0_D$'' indicates the Dalitz mode of neutral pion decay, $\pi^0 \to \gamma e^+ e^-$. By unfolding the number of decays observed in this mode ($7.38 \times 10^5$) with the $K_L \to \pi^+ \pi^- \pi^0_D$ branching ratio, we estimate the total number of parent $K_L$ decays in the selection sample, around $5 \times 10^8$. Assuming that the selection efficiency of this sample for the Dalitz mode and the $K_L \to \pi^+ \pi^- e^+ e^-$ mode (and our signal) are similar, we then perform a binned $\chi^2$ analysis of the $m_{ee}$ spectrum as before to extract a limit on the branching ratio for $K_L \to \pi^+ \pi^- X_{ee}$. We estimate a bound of
\begin{align}
    \mathcal{B}(K_L &\to \pi^+ \pi^- X_{ee}^{17}) < 2.18 \times 10^{-8} \, ,
\end{align}
which can be interpreted as a bound on the effective coupling combination in \cref{eq:KLPiPlusMinus_couplings}.

By scanning over the $m_{ee}$ data for both decay modes, we report the exclusions over the effective coupling combinations as a function of $m_X$ in \cref{fig:k_limits_by_mass}.

\subsection{$K \to \pi \pi X_{\rm inv}$ decays}
Searches for an invisible resonance have been performed in the three-body decays of kaons.
The E787 experiment sets limits as strong as 
\begin{equation}
    \mathcal{B}(K^+ \to \pi^+ \pi^0 \nu \overline \nu) \lesssim 4.3 \times 10^{-5} \text{ at 90\% CL~\cite{E787:2000iwe}},
\end{equation}
which constrains light invisible $X$ bosons as well.
For the $K_L$, the constraint is stronger,
\begin{equation}
    \mathcal{B}(K_L \to \pi^0 \pi^0 \nu \overline \nu) \lesssim 8.1 \times 10^{-7} \text{ at 90\% CL~\cite{E391a:2011aa}}
\end{equation}
with corresponding constraints on the case of $K_L \to \pi^0 \pi^0  X_{\rm inv}$.

Such three-body decays have also been discussed in the context of ALPs and the QCD axion of ~\cite{Alves:2020xhf}.
It has also been noted that pion reinteractions can significantly enhance analogous branching ratios in the $\eta$ and $\eta'$ cases~\cite{Alves:2024dpa}.

The decays $K_{S,L} \to \pi^+ \pi^- X_{\rm inv}$ have been discussed in the context of massless dark photons in Refs.~\cite{Fabbrichesi:2017vma,Su:2020xwt}.
There, the decay proceeds via dipole-transitions of the type $X^{\mu \nu} \left(\overline s \sigma_{\mu \nu} d \right)$.
Because the dark photon is massless, angular momentum and gauge invariance forbid direct $K\to \pi X$ transitions that would otherwise provide the leading constraints.

\subsection{$K^+ \to \pi^+ \gamma X_{ee}$ decays}

A closely related process is the radiative $K\to\pi \gamma X$, which, as we will see, at $\mathcal{O}(p^2)$ only exists for the charged kaon mode.
The analogous processes $K^+ \to \pi^+ \gamma \gamma$ and $K_L \to \pi^0 \gamma \gamma$ have long been studied and are known to vanish at $\mathcal{O}(p^2)$ ~\cite{Ecker:1987hd}, appearing only at loop-level at $\mathcal{O}(p^4)$~\cite{Ecker:1987fm,Cappiello:1988yg}.
This is because for the photon, it can only proceed via gauge-invariant operators proportional to $F_{\mu\nu}F^{\mu\nu}$, demanding a higher number of derivatives.
This argument, however, is not applicable to vectors coupled to non-conserved currents, and so the replacement of a photon by $X$ can induce the decay at leading order, even for the mixed case involving both $\gamma$ and $X$.
The case of double $X$ emission is treated in \cref{sec:KtopiXX}.

With $X\to e^+e^-$, we  can take advantage of the experimental observation of the rare SM process $K^+\to \pi^+\gamma e^+e^-$ by NA48/2~\cite{NA482:2007kxb}.
The measured branching ratio is
\begin{align}
    \mathcal{B}(K^+ \to \pi^+ \gamma e^+e^-) = (1.19 \pm 0.13)\times 10^{-8}.
\end{align}
No cut on the invariant mass of $m_{ee}$ was reported, but the above result is only valid for $m_{\gamma ee} > 260$~MeV, where the bulk of SM events are.
In our case, this suggests that $m_{\gamma X}^2 = (p_X +p_{ee})^2= m_X^2 + 2 (p_{ee} \cdot p_\gamma)> (260$~MeV$)^2$, which for very light vectors, including \xseventeen, is approximately just a requirement on the kaon-rest-frame angle between $X$ and $\gamma$, namely 
$m_{\gamma ee}^2 \simeq 2 E_X E_\gamma(1-\cos\theta_{\gamma,X})$, meaning that the signal selection will prefer $\gamma X$ final states well-separated in angle.
As we will see, this is precisely what is preferred for the emission of an $X$ coupled to a non-conserved current.

In the diagonal basis for the meson kinetic terms, we find that 3 different diagrams contribute to $K^+\to \pi^+ \gamma X$ as shown in \cref{fig:KtoPiXgamma_diagram}.
In the non-diagonal basis, a total of 13 diagrams would contribute at the same order.
We do not include the contribution from intermediate $\pi^0,\eta,\eta' \to\gamma X$ as we expect these diagrams to be subdominant for $m_{\gamma ee} > 260$~MeV.
The total amplitude is given by $\mathcal{M} = (\epsilon^\gamma_\mu \epsilon^X_\nu)^*  \mathcal{M}^{\mu \nu}$, where $\mathcal{M}^{\mu \nu} = \sum_{i} \mathcal{M}^{\mu \nu}_i$ is the total amplitude expressed in terms of
\begin{align}
    \mathcal{M}_{1}^{\mu \nu} &= 2e \kappa_8   g_{K\pi^+}^X \,g^{\mu \nu}
    \\
    \mathcal{M}_{2}^{\mu \nu} &= 4e \kappa_8 g_{K\pi^+}^X \left[\frac{p_K^\mu p_\pi^\nu}{2p_K\cdot k_\gamma} -  \frac{p_\pi^\mu p_K^\nu}{2p_\pi\cdot k_\gamma}\right]
\end{align}
One can check that for a conserved current with flavor-universal couplings, one gets $g_{T_3}^V = g_{V_3}^V$ (i.e., $g_d^V =g_{s}^V$), canceling the amplitudes exactly at this order.
In fact, the conservation of the EM current is explicitly manifest in
\begin{align}
    k_{\gamma}^\mu (\epsilon_X^{\nu*}\mathcal{M}_{\mu \nu}) &= -2 eF^2G_8 g_{K\pi^+}^X (\epsilon_X^* \cdot k_{X}) = 0,
\end{align}
with $g_{K\pi^+}^X$ as defined in \cref{eq:g_kpi}.
Therefore, from the above we can conclude that the total decay rate must then be proportional to $g_{K\pi^+}^X$.

\begin{figure*}[t]
    \centering
    \includegraphics[width=1.0\linewidth]{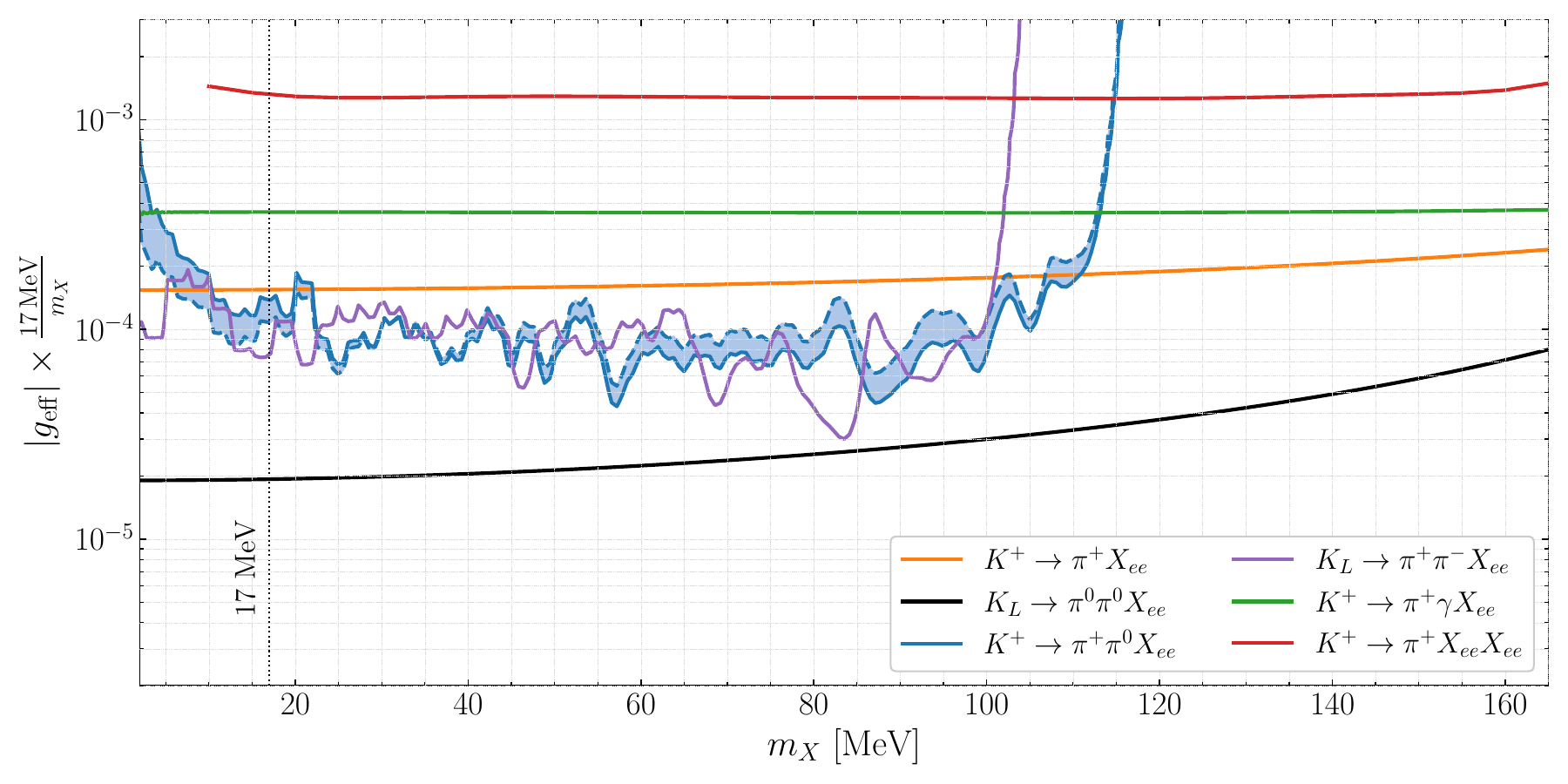}
    \caption{Limits derived from kaon decay data on promptly-decaying vector bosons $X$ as a function of $m_X$ and the effective coupling $g_{\rm eff}$. For all channels except $K^+ \to \pi^+ X$ and $K^+ \to \pi^+ \gamma X$, both proportional to $g_{K\pi^+}$ defined in \cref{eq:g_kpi}, we only take couplings proportional to contact diagrams to contribute such that a single combination factorizes to $g_{\rm eff}$; see Table~\ref{tab:constraints}. For $K^+ \to \pi^0 \pi^+ X_{ee}$ decays (blue), the band envelops limits across two decay model assumptions: ``IB'' (solid) or ``M'' (dashed). For $K^+ \to \pi^+ X_{ee} X_{ee}$, there are no direct limits below $m_X = 10$ MeV due to the NA62 selection.
    }
    \label{fig:k_limits_by_mass}
\end{figure*}

The full decay rate is rather cumbersome, so we expand in $r_X = (m_X/m_K)^2$ (but not in $r=(m_\pi/m_K)^2$), keeping all masses in the propagators and using the dimensionless invariant masses $x_{ij} = m_{ij}^2/m_K^2$ to find
\begin{equation}
    \frac{\dd \Gamma_{K^+ \to \pi^+ \gamma X}}{\dd x_{\gamma X} \dd x_{\pi X}} = \frac{\alpha g_{K\pi^+}^2}{16 \pi^2} \frac{G_8^2 F^4 m_K^3}{m_X^2} x_{\gamma X}\frac{N_0}{D_1^2 D_2^2}.
\end{equation}
where $N_0=x_{\pi \gamma} x_{\pi X} - r(1-x_{\pi \gamma}(1-2x_{\pi X}))$, $D_1 = r-x_{\pi \gamma}$, and $D_2 = (r + r_X - x_{\pi \gamma} - x_{\gamma X})$.
The proportionality to $x_{\gamma X}=m_{\gamma ee}^2/m_K^2$ shows that the experimental cuts on $m_{\gamma ee}^2$ do not preclude us from setting a competitive limit. 
From NA48/2 data on $m_{\gamma e e}$~\cite{NA482:2007kxb}, we set a limit on the coupling $g_{K\pi^+}$ shown in \cref{fig:k_limits_by_mass} (green), comparable to $K^+ \to \pi^+ X$ limits (orange).
We emphasize that improvements to both of these channels are possible with dedicated bump hunts on top of the SM backgrounds by experimental collaborations.

\subsection{$K^+ \to \pi^+ X_{ee} X_{ee}$ decays}
\label{sec:KtopiXX}

Due to the significant longitudinal mode enhancement to kaon decays, one can also obtain new limits from the double emission of vector particles.
Of special relevance is double-bump hunt in the five-charged track final state of charged kaon decays at NA62~\cite{NA62:2023rvm}, which sets a limit on the branching of $K^+ \to \pi^+ X_{17} X_{17}$ of
\begin{equation}
    \mathcal{B}(K^+ \to \pi^+ X_{ee} X_{ee})\big\vert_{m_X = 17 \, \text{MeV}} < 3.1\times10^{-9} \,\text{ at 90\% CL}.
\end{equation}
The KOTO experiment has also performed a preliminary search for the analogous channel in the neutral mode~\cite{Li:2024ieq},
\begin{equation}
    \mathcal{B}(K_L \to \pi^0 X_{ee} X_{ee})\big\vert_{m_X = 100 \, \text{MeV}} < 4.11\times10^{-7} \,\text{ at 90\% CL}.
\end{equation}
A future KOTO analysis may be able to improve upon this limit and extend it to lower $X$ masses.
At present, we can only use the constraint in the charged mode to set limits on \xseventeen.

The decay $K^+(p_K) \to \pi^+(p_\pi) + X_\mu (k_1) + X_\mu (k_2)$ can proceed via three separate amplitudes, analogous to the ones shown in \cref{fig:KtoPiXgamma_diagram}. 
These are given by
\begin{widetext}
\begin{align}
    \mathcal{M}_{1}^{\mu \nu} &= 4 \kappa_8  g^{\mu \nu} g_{K \pi^+}^{XX}
    \\
    \mathcal{M}_{2}^{\mu \nu} &= 4 \kappa_8 g_{K \pi^+}^{X}  \left[ g_{V_3}^V\frac{p_K^\mu p_\pi^\nu}{2p_K\cdot k_1 - m_X^2} -  g_{T_3}^V\frac{p_\pi^\mu p_K^\nu}{2p_\pi\cdot k_1 + m_X^2}\right] 
    \\
    \mathcal{M}_{3}^{\mu \nu} &= -4 \kappa_8 g_{K \pi^+}^{X}  \left[ g_{V_3}^V\frac{p_K^\mu p_\pi^\nu}{2p_\pi\cdot k_2 + m_X^2} -  g_{T_3}^V\frac{p_\pi^\mu p_K^\nu}{2p_K\cdot k_2 - m_X^2}\right] 
\end{align}
where we defined the coupling
\begin{align}
  \label{eq:g_kpiXX}
    g_{K \pi^+}^{XX} &\equiv \left(
      g^V_{T_3}g^V_{V_3} 
    + g^V_{T_3}g^A_{T_0} 
    + g^V_{V_3}g^A_{V_0} 
    + g^V_{T_0}g^A_{V_0} 
    - (g^A_{U_0})^2\right)
    -4\frac{g^L_{T_3}g^R_{T_3}+g^L_{V_3}g^R_{V_3} r}{1-r}
\end{align}
To leading order in powers of $m_K/m_X$, the differential decay rate reads
\begin{equation}
\frac{d \Gamma}{dx_{12} dx_{\pi 1}} = \frac{\kappa_8^2}{512 \pi^3}\frac{m_K^5}{m_X^4} \left(g_{K\pi^+}(g_{V_3}^V - g_{T_3}^V)(1-2x_{\pi 1}+r) + 2 x_{12} (g_{K\pi^+}^{XX} - g_{V_3}^Vg_{T_3}^V) \right)^2.
\end{equation}
\end{widetext}
where we account for identical $X_\mu$ particles in the final state.

\begin{figure*}[t]
    \centering
    \includegraphics[width=0.49\linewidth]{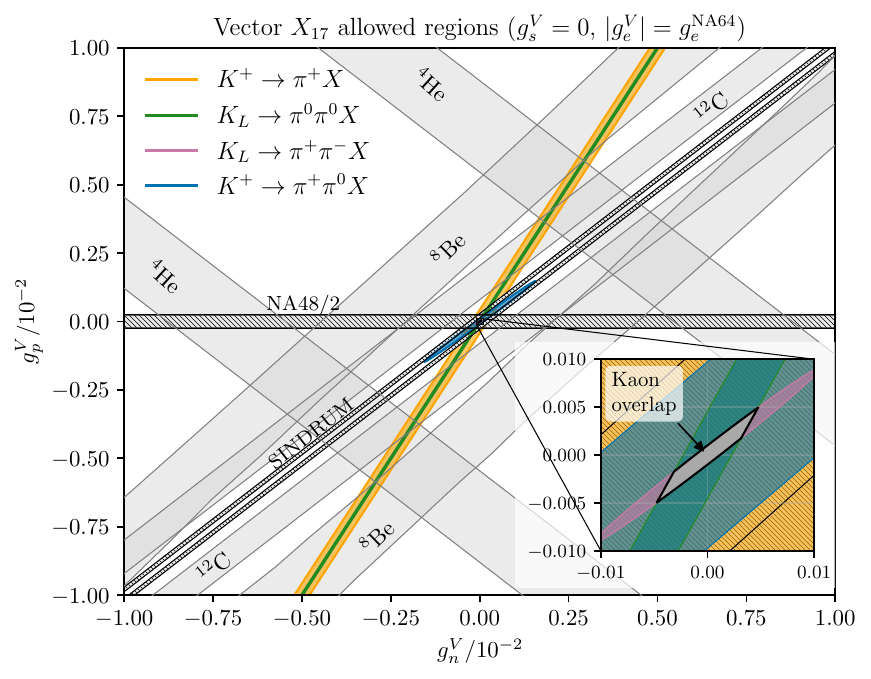}
    \includegraphics[width=0.49\linewidth]{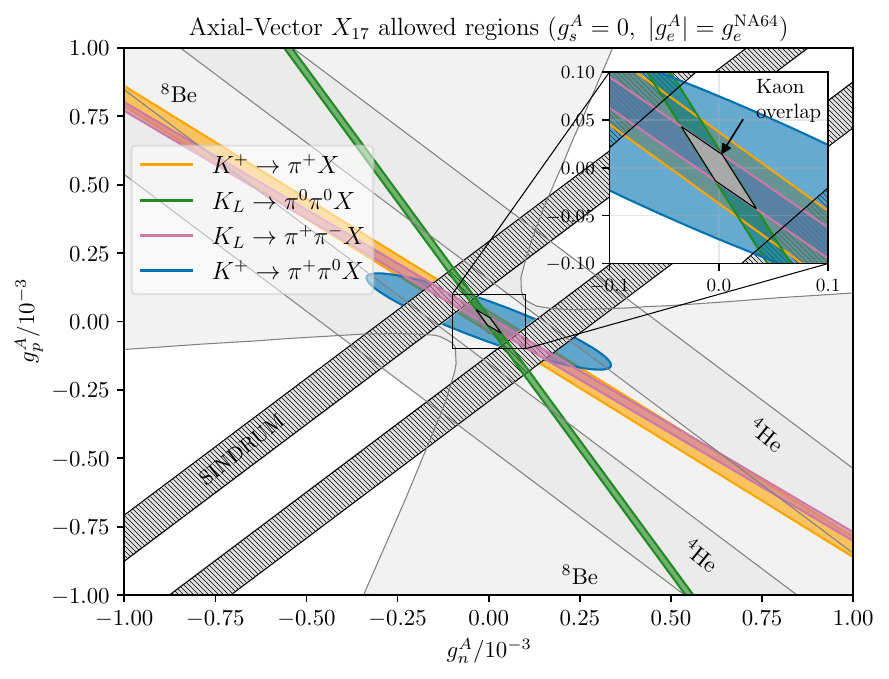}
    \caption{The allowed regions in the 17 MeV spin-1 boson coupled to a vector (left) or axial-vector (right) current for $g_{s}^{V,A} = 0$, setting the electron coupling to saturate the lower bound from NA64~\cite{NA64:2019auh}. 
    The $90\%$ CL allowed regions for different kaon decay channels are shown in color, together with their overlap, which is highlighted by the inset axis.
    Constraints from pion decays are shown as dark grey bands. 
    For SINDRUM, the two bands correspond to different sign choices for $g_e^{V}$ (left) or $g_e^{V}$ (right).
    The $2\sigma$ allowed regions for the $^4$He, $^8$Be, $^{12}$C ATOMKI anomalies are shown as light shaded grey (from \cite{BarducciErratum}).
    The preference regions for $^{12}$C in the axial-vector case are beyond the axis boundaries~\cite{Mommers:2024qzy}.
    }
    \label{fig:2D_parameter_gs_zero}
\end{figure*}

A similar calculation can be done for the neutral channel involving $K_L$.
In that case, the amplitude for the contact term is proportional to
\begin{equation}
    \label{eq:g_kpiXXneutral}
    g_{K\pi^0}^{XX} \equiv g^A_{U_0}g^R_{U_3} + (g^L_{U_0}g^R_{U_0})\frac{2r}{1-r}.
\end{equation}
No IB diagram exists with emission from the pion line, so amplitudes $\mathcal{M}_{2,3}$ are both proportional to $g_{K\pi^0}^X g_{U_3}^V$.
We find that the eta pole does not contribute to this decay at $\mathcal{O}(p^2)$.

We make a comparison of the constraints from all the channels discussed in this section in \cref{fig:k_limits_by_mass}; from least to most stringent, these are $K^+ \to \pi^+ X_{ee} X_{ee}$ (red), $K^+ \to \pi^+ \gamma X_{ee}$ (green), $K_L \to \pi^+ \pi^- X_{ee}$ (purple), $K^+ \to \pi^+ X_{ee}$ (orange), $K^+ \to \pi^+ \pi^0 X_{ee}$ (blue), and $K_L \to \pi^0 \pi^0 X_{ee}$ (black). Since all channels except for $K^+ \to \pi^+ X_{ee}$ and $K^+ \to \pi^+ \gamma X_{ee}$ are not factorizable into a single coupling combination independent of $m_X$, for Fig.~\ref{fig:k_limits_by_mass} we make a choice to set $U_3$, $T_3$, and $V_3$ couplings to zero; in this limit, only the contact diagrams remain and amplitudes can be expressed in terms of a single, factorizable effective coupling we call $g_{\rm eff}$. See Table~\ref{tab:constraints} for a breakdown of how each channel depends on various coupling combinations with interference terms, and the $g_{\rm eff}$ taken (second-to-last column).

\section{Results for \xseventeen}
\label{sec:results}

We now discuss our constraints in the context of \xseventeen and the ATOMKI anomalies.
We consider \xseventeen coupled exclusively to either vector or axial-vector currents as chiral currents are strongly constrained by atomic parity violation.
For vector couplings, it will be convenient to define the nucleon couplings $g^V_p = 2g^V_u + g_d^V$ and $g^V_n = 2 g^V_d + g_u^V$.
For axial-vector couplings, we use the matrix elements $\kappa^{u} = 0.817$ and $\kappa^{d} = −0.450$ from \cite{Alexandrou:2019brg} to define $g_p^A = \kappa^{u} g_u^A + \kappa^{d} g_d^A$ and $g_n^A = \kappa^{d} g_u^A + \kappa^{u} g_d^A$.
We neglect the $\mathcal{O}(10\%)$ effect the strange quark contribution to the nucleon couplings.

Our bounds ought to be compared with those from pion decay, including the $\pi^0 \to \gamma X_{ee}$ limit from NA48/2~\cite{NA482:2015wmo,Feng:2016jff},
\begin{equation}
    \left|2g_u^V + g_d^V\right| < 8.0 \times 10^{-4} \text{  at 90\% CL}
\end{equation}
and the $\pi^+ \to e^+ \nu_e X_{ee}$ limit from SINDRUM~\cite{SINDRUM:1989qan,Hostert:2023tkg},
\begin{equation}
    \left|g_u^R-g_d^R - g_\nu^L + g_e^L\right| < 8.5 \times 10^{-5} \text{  at 90\% CL}.
\end{equation}
Direct constraints on the coupling to electrons include the limit by the KLOE $e^+e^-$ collider~\cite{Anastasi:2015qla}, which sets an upper bound of $\sqrt{(g_e^V)^2 + (g_e^A)^2} = e \varepsilon < 6 \times 10^{-4}$ while the constraints from NA64~\cite{NA64:2018lsq,NA64:2019auh,NA64:2021aiq} provide a lower limit of $\sqrt{(g_e^V)^2 + (g_e^A)^2} > g_e^{\rm NA64} = 2.1\times 10^{-4}$.
The lower limit from NA64 is more stringent than the lifetime requirement on \xseventeen for it to explain the ATOMKI results.
Constraints on the coupling of \xseventeen to neutrinos are discussed in \cite{Denton:2023gat}, but we do not consider these due to the stringent limits from neutrino scattering data.

\cref{fig:2D_parameter_gs_zero} shows the allowed regions in the two-dimensional parameter space of proton and neutron couplings for $g_{s}^{V,A}=0$.
The $2\sigma$ ATOMKI preferred regions are shown in the plot as grey bands and the pion decay limits are shown as hatched grey bands.
The SINDRUM constraint is shown when saturating the lower limit on $g_e^{V,A}$ from NA64.
Increasing the value of $g_e^{V,A}$ (for example by using values preferred by the PADME experiment~\cite{PADME:2025dla} if interpreted as a positive signal of $X_{17}$), will increase the tension between kaon and pion results further.

We can see that in the vector \xseventeen model there is no overlap between the ATOMKI regions and the kaon decay limits.
The latter complement the pion decay bounds that already excluded the vector interpretation of the ATOMKI anomalies.
In the axial-vector case, the couplings required to explain the $^8$Be results from ATOMKI can be extremely small according to the fit in \cite{BarducciErratum}, so it may be possible to reconcile it with kaon limits.
However, a simultaneous $^{4}$He and the $^{12}$C explanation requires significantly larger couplings and, therefore, are in severe tension with the regions allowed by kaon decays.
Indeed, the $^{12}$C fit from \cite{Mommers:2024qzy} does not even appear within the scale of our plot.

\begin{figure*}
 \includegraphics[width=0.49\linewidth]{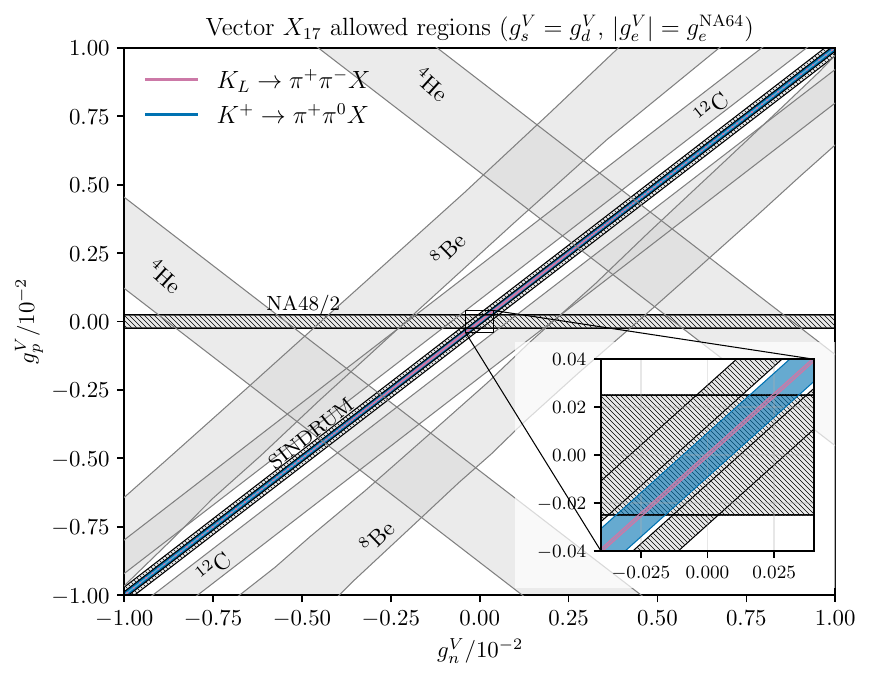}
    \includegraphics[width=0.49\linewidth]{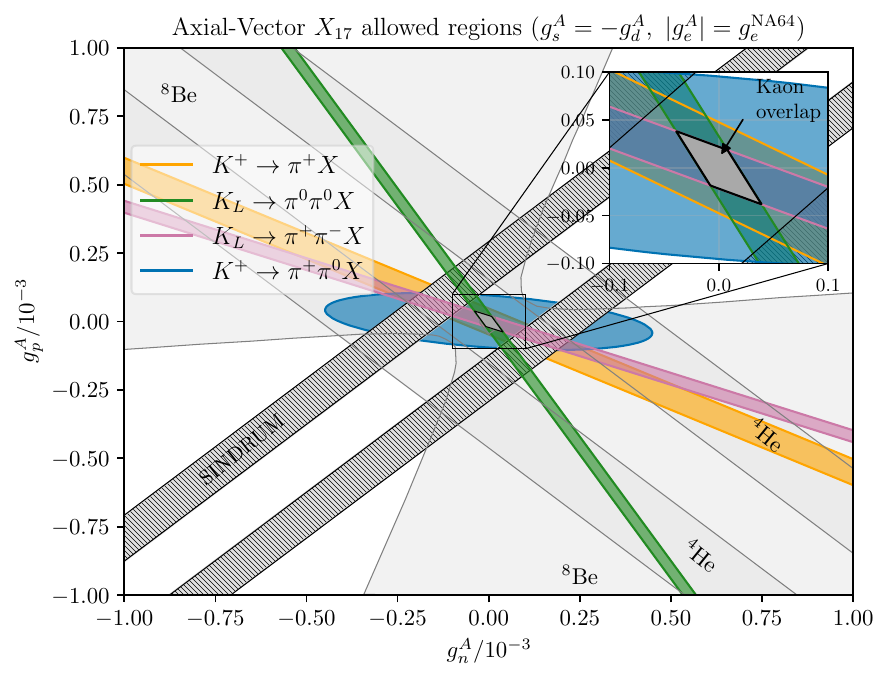}
    \caption{Same as \cref{fig:2D_parameter_gs_zero}, but choosing $g_d^V = g_s^V$ (left) and $g_d^A = -g_s^A$ (right).
    In the vector case, this corresponds to a $U$-spin symmetric scenario, where the vector cannot be radiated off of neutral kaons.
    \label{fig:2D_parameter_gs_canceled}
    }
\end{figure*}

Moving away from the limit of $g_s^{V,A} = 0$, one can ask whether it is possible to relax the kaon decay constraints by fine tuning $g_s^{V,A}$ so as to enlarge the allowed region in \cref{fig:2D_parameter_gs_zero}.
One scenario is shown in \cref{fig:2D_parameter_gs_canceled}, where in the vector model we impose $U$-spin conservation. 
This shuts off emission from neutral kaons, though the vector can still be emitted from charged pion legs of our diagrams when $g_u^V \neq g_d^V$.
The fact that the SINDRUM limits are much more sensitive to the choice of $g_e^V$ than the kaon ones explains why the diagonal bands from SINDRUM do not overlap with those from kaon decays involving charged pions.
In the axial-vector case we repeat a similar exercise but for $g_u^A = - g_s^A$.

\begin{figure*}[t]
    \centering
    \includegraphics[width=0.49\linewidth]{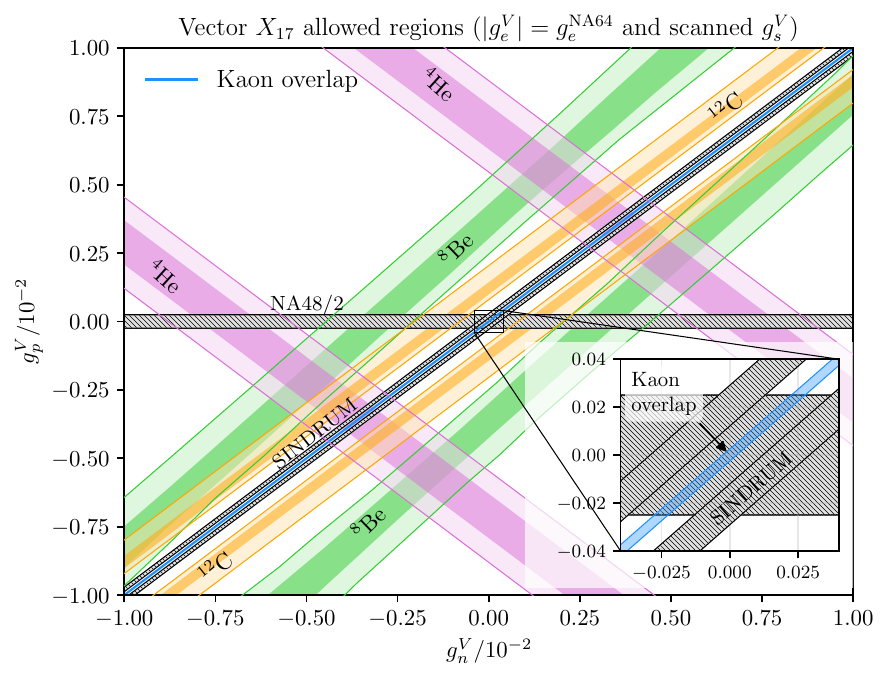}
    \includegraphics[width=0.49\linewidth]{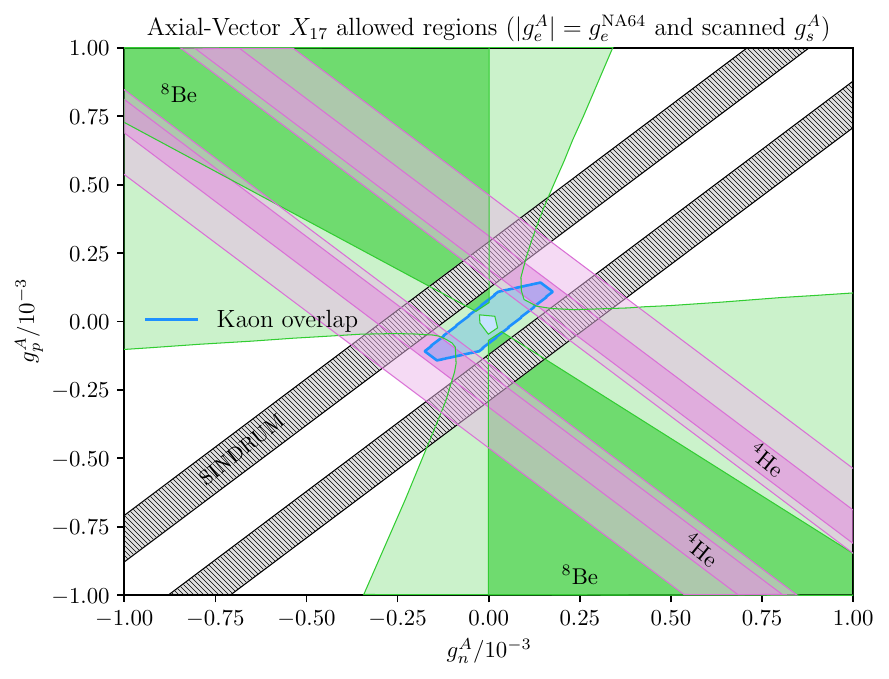}
    \caption{The kaon decay constraints on the 17 MeV vector (left) and axial-vector (right) bosons. 
    For each point in the 2D axis, we scan $g_s^{V}$ (left) or $g_s^{A}$ (right) values to find a region allowed by all kaon limits considered in \cref{fig:2D_parameter_gs_zero}.
    The misalignment of regions preferred by the $^4$He, $^8$Be, $^{12}$C anomalies with the kaon decay constraints permits no simultaneous union of parameter space to consistently explain the anomalies in this scenario.
    In the vector scenario, the allowed band corresponds to an $X$ boson that couples to an SU$(3)$-flavor symmetric current ($g_u^V = g_d^V = g_s^V$).
    In this case, the vector just couples to the baryonic current and cannot be radiated off in meson decay.}
    \label{fig:2D_parameter_gs_scanned}
\end{figure*}

In \cref{fig:2D_parameter_gs_scanned}, we go one step further and scan the strange quark coupling~ in search for a region where all kaon decay bounds can be satisfied for a given choice of $g_s^{V,A}$.\footnote{We refer to this method as ``scanning $g_s^{V,A}$" and note that it is not the same as profiling since we do not study the likelihood function for any of the observables considered here.}
Because of the enhanced freedom, a larger allowed region appears.
In the vector boson case, this region corresponds to the SU$(3)$-symmetric limit, where $X$ couples to the baryonic current and its emission in kaon decays (as calculated in this article) vanishes.
Even in this limit, however, other constraints from anomaly diagrams may apply.

\section{Opportunities with $\pi^-$ capture}
\label{sec:pions}

An experimental situation with ATOMKI excess interpreted as $X_{17}$ is still unsettled. 
While claimed in a variety of a few nucleon systems, the entire $X_{17}$ hypothesis rests on the results by one group, and fully independent confirmation or sufficiently precise null results by other groups are needed. 
Moreover, so far the effect is claimed in nuclear emissions that are mostly associated with formation of $\alpha$-particles inside light nuclei. While it is hard to mimic the observed angular structure of $e^+e^-$ pairs with nuclear physics without making almost arbitrary assumptions about nuclear form factors \cite{Zhang:2017zap}, it makes sense to investigate the presence/absence of $X_{17}$ in even simpler systems. 
Repeating the arguments of Ref. \cite{Chen:2019ivz}, systems with one or two nucleons are ideal as they can be fully treated from first principles, and to that end one could search for $X_{17}$ and other light particles in the $\pi^-$ capture on protons and deuterons. 

Capture of $\pi^-$ is a well studied process where the pion forms an atomic system with a nucleus and is subsequently absorbed. An exotic particle emission proceeds via
\begin{equation}
    \pi^- + (A,Z)  \to X_{17} + (A, Z-1).
\end{equation}
For hydrogen, the capture process produces an approximately mono-energetic $X$ particle that subsequently decays to $e^+e^-$. Such a test would bring different kinematics in the angular distribution of the $X_{17}$ decay pair. Given $O(130)$\,MeV energy release absorbed mostly by $X$, the angular feature due to $X_{17}$ is shifted to $\sim 16$ degrees in the opening angle \cite{Chen:2019ivz}.  
Note that the new physics will compete with the charge-exchange reaction $\pi^- + p^+ \to \pi^0_D + n$ with $\pi^0_D \to \gamma e^+e^-$ a pion Dalitz decay.
The same reaction on a deuterium target, $\pi^- + d \to \pi^0 + 2n$, is strongly suppressed by kinematics and angular momentum conservation, and so while $X_{17}$ production on a deuterium target may be a factor of $\sim 2$ smaller than on a Hydrogen target, it would not suffer from the same $\pi^0_D$ backgrounds.

Long ago the $d(\pi^-,e^+e^-)nn$ reaction was already studied, including the distribution over the invariant mass of the pair \cite{kloepppel1964interactions}. A ``bump hunt" over this distribution, when measured with higher statistics, will reveal the absence/presence of $X_{17}$. We believe that one should repeat these studies, possibly at PSI, where negative pion capture on protons is  used by the MEG-II collaboration for the calibration of the photon calorimeter of their detector \cite{MEGII:2018kmf}.

\section{Discussion and Conclusions}
\label{sec:conclusions}

In this work, we have provided an analysis, channel-by-channel, of precision measurements on kaon decays with a hypothetical light (Lorentz-) vector particle $X_\mu$ in the final state, followed by the prompt decay of $X$ to an electron-positron pair. 
If $X_\mu$ couples to vector or axial-vector currents of light quarks and is kinematically accessible, it would alter the observed rare kaon decays with pion and $e^+ e^-$ final states. 
We make use of the $\Delta S = 1$ violating operators in the chiral effective field theory, parameterized by an octet coupling $G_8$ that is known to dominate many kaon decay amplitudes. 
General rules of ChPT are then applied to include the $X$ couplings at the level of the meson interactions.  
While all couplings of $X$ particles to quarks are assumed to be small, many decay channels exhibit longitudinal enhancement of amplitudes with outgoing light $X$ bosons. 
Moreover, because we do not impose that $X$ couples to a conserved current, the new decay modes can circumvent SM selection rules and dominate over kaon decay modes that are strongly suppressed in the SM that would otherwise provide a background.
A prime example of this is the $K_L\to \pi^0\pi^0e^+e^-$ decay mode, extremely suppressed in the SM by the small CP-violating amplitude and the absence of an internal bremsstrahlung amplitude.
If $X$ couples to a non-conserved current, the decay $K_L\to \pi^0 \pi^0 X$ can proceed to an on-shell longitudinal $X$, making it a CP-conserving channel that can easily dominate the corresponding SM contribution for couplings as small as $g_{u,d,s}^{V,A} \sim \mathcal{O}(10^{-5})$.

The sensitivity of the kaon decays to quark couplings of $X$-particle adds another obstacle for a tentative explanation of nuclear anomalies by an $X_{17}$. 
Predictive models with minimum number of floating parameters, such as the axion model of Ref.~\cite{Alves:2017avw,Alves:2020xhf}, is already excluded by rare kaon decays~\cite{NA62:2023rvm}. 
In case of Lorentz-vector $X_{17}$, a purely phenomenological approach is sometimes taken that would typically assume a collection of either $g_u^V,g_d^V$ or $g_u^A,g_d^A$ couplings that satisfy $X_{17}$ requirements in nuclear experiments, yet somehow escaping other constraints from the fixed target experiments and pion decays. (Having both vector and axial-vector couplings at the same time would break spatial parity at the level that would exceed SM weak breaking, $g_{u,d}^Ag_{u,d}^V(17\,{\rm MeV})^{-2}\gg G_F$, which could be incompatible with, {\em e.g.}, the results of Ref.~\cite{Wood:1997zq}.) 
These two-parameter models of $X_{17}$ are already in great tension with the existing data, including an internal tension between the axial-vector couplings preferred by the $^{12}$C, $^4$He, and $^8$Be ATOMKI results~\cite{Mommers:2024qzy}.

When adding a strange quark into the mix, we enlarge the parameter space of couplings by including $g_s^{V,A}$, but at the same time the space of sensitive observables is expanded much more by including all the modes discussed in this paper. 
We find that the combination of kaon bounds strongly constrains the parameter space of such models since the different decay channels considered here are sensitive to a variety of linear combinations of up, down, and strange quark.
Indeed, different modes are sensitive to different couplings in the singlet and triplet combinations of the flavor subgroups of the $SU(3)$ chiral perturbation theory (the $T$, $U$, and $V$ operators, i.e., isospin, $U$-, and $V$-spin subgroups). 
Existing limits from the non-observation of $X_{17}$-like excesses in kaon data leave no room for an explanation with vector couplings and little room for an axial-vector coupling scenario.
While a joint statistical fit would be needed to establish the degree of exclusion or tension in these models, our work shows that the 90\% CL excluded regions by kaon data cover most or all of the regions preferred by the ATOMKI $^{12}$C, $^4$He, or $^8$Be data.

There are several experimental efforts designed to test the ATOMKI-derived $X_{17}$ hypothesis. 
We will briefly touch upon some of them that recently reported results. An experiment in Vietnam~\cite{Anh:2024req} claims to observe deviations in $e^+e^-$ pair-production in decays of excited $^8$Be state consistent with $X_{17}$ hypothesis. On the other hand, MEG-II collaboration reported no evidence of angular anomalies in the same channel~\cite{MEGII:2024urz}.
Recently an intriguing measurement by the PADME experiment~\cite{PADME:2025dla} detected a $2$-to-$3\,\sigma$ excess of the electron-positron pairs with an invariant mass of $\simeq 17$\,MeV, which perhaps lends some support to the $X_{17}$ hypothesis. 
This search is conducted via scattering of energetic positrons on fixed target and is not sensitive to the quark couplings. A minimal explanation of \cite{PADME:2025dla} in terms of a new particle, marginally consistent with $\pi^0$ decay constraints, would point, in fact, to a dark photon, {\em i.e.}, $X_{17}$ coupled to a conserved current, thereby evading most of the kaon and charged pion decay constraints. The strength of the dark photon coupling consistent with \cite{PADME:2025dla} would, however, be significantly smaller than that inferred from the ATOMKI results. In a purely speculative vein, one could imagine that interpretations of $X_{17}$ in nuclear transitions are in fact affected by some unidentified backgrounds, and actual signal excess is smaller than claimed, and hence consistent with the dark photon interpretation. While such an explanation is also quite implausible, it is perhaps more minimal than extremely tuned and theoretically very poorly motivated scenarios of vectorial $X_{17}$ with fine-tuned couplings to quarks. 
This is why it is important to achieve independent experimental scrutiny and verification of $X_{17}$ with hadronic systems.
The negative pion capture method advocated here in \cref{sec:pions} and in Ref.~\cite{Chen:2019ivz} is one possible way to achieve progress. 

\vspace{0.5em}

\paragraph*{Note Added:}
During the completion of this work, Ref.~\cite{Fieg:2026zkg} appeared, where the authors study spin-1 bosons coupled to chiral SM currents. 
One of their findings is that the $^{12}$C results of ATOMKI drives the tension with existing experiments, including the pion decay constraints of Ref.~\cite{Hostert:2023tkg}. 
They also point out that without $^{12}$C, there are regions of the parameter space that provide viable explanations of the ATOMKI $^8$Be and $^4$He results provided that fine-tuning between $g_e^V$ and chiral quark couplings is invoked to relax pion decay limits. 
Our results further constrain this fine-tuned solution since, irrespective of the strange coupling, there is still tension between ATOMKI and kaon results.
We echo the point made in \cite{Fieg:2026zkg} that a more rigorous statistical statement of the tension between ATOMKI and other constraints has not been quantified, including the internal tension between different ATOMKI results.

\paragraph*{Acknowledgements:} 
We would like to thank Evgueni Goudzovski, David McKeen, and Angela Papa for discussions on this topic.
This work was partially supported by the University of Iowa’s Year 2 P3 Strategic Initiatives Program through funding received for the project entitled ``High Impact Hiring Initiative (HIHI): A Program to Strategically Recruit and Retain Talented Faculty.'' The work of AT is supported in part by the U.S. DOE grant \#DE-SC0010143. 
The work of MP is supported in part by U.S. Department of Energy Grant \#DE-SC0011842.


\appendix 

\section{$X$ in U$(3)$ ChPT}
\label{app:chpt}

We work with the U$(3)$ ChPT, adding the $X$ boson as an external field. 
Unless otherwise specified, we impose the on-shell condition $\partial_\mu X^\mu= 0$.
Our basis is defined as $U = e^{i\Phi/F}$, with $F \simeq 93$~MeV and the goldstone fields
\begin{align}
    & \Phi = \frac{1}{\sqrt 2}\sum_{a=0}^8 \phi_a\lambda_a 
    \\\nonumber 
        &= \left( 
    \begin{matrix}
\pi^0 +  \frac{\eta_8}{\sqrt{3}} + \phi_0 & \sqrt{2} \pi^+ & \sqrt{2} K^+ \\
\sqrt{2} \pi^- & -\pi^0 + \frac{\eta_8}{\sqrt{3}} + \phi_0 & K_L-i K_S \\
\sqrt{2} K^- & K_L+i K_S & -\frac{2\eta_8}{\sqrt{3}} + \phi_0
\end{matrix}
    \right)
\\\nonumber
    &= \pi^0 \lambda_3 + \eta_8 \lambda_8 +  \eta_0 \lambda_0 
    \\\nonumber
    &\quad\quad  + \sqrt{2} \left(\pi^+T_+ + K^+V_+ + K^0U_+ + \text{h.c.}\right),
\end{align}
where in addition to the usual SU$(3)$ octet mesons we added the chiral singlet $\phi_0 = (\sqrt{2/3})\eta_0$ with the identity matrix $\lambda_0= (\sqrt{2/3})\text{diag}(1,1,1)$.

We neglect CP-violation in the SM throughout this article, so we can identify $K_S$ with the kaon that is a CP-even eigenstate. 
This field, therefore, carries the imaginary unit that distinguishes it from the CP-odd mesons:
\begin{equation}
    K^0 = \frac{K_L -i K_S}{\sqrt{2}}, \,\, \bar{K^0} = \frac{K_L + i K_S}{\sqrt{2}}.
\end{equation}
When convenient, we also make use of the three SU(2) flavor subgroups, namely $T$-spin (isospin), $U$-spin and $V$-spin.
In particular, we define the following
\begin{align}
    T_0 &= \text{diag}(1,1,0), \, T_\pm = \frac{\lambda_1 \pm i\lambda_2}{2}, \, T_3 = \lambda_3,
    \\\nonumber
    U_0 &= \text{diag}(0,1,1), \, U_\pm = \frac{\lambda_6 \pm i\lambda_7}{2}, U_3 = \frac{\sqrt{3}\lambda_8 - \lambda_3}{2},
    \\\nonumber
    V_0 &= \text{diag}(1,0,1), \, V_\pm = \frac{\lambda_4 \pm i\lambda_5}{2}, V_3 =  \frac{\sqrt{3}\lambda_8 + \lambda_3}{2}.
\end{align}

At $\mathcal{O}(p^2)$, the relevant terms in the Lagrangian are given by
\begin{equation}
    \mathcal{L}_{p^2} = \frac{F ^2}{4} \langle (D_\mu U)^\dagger D^\mu U\rangle + v \langle M U + U^\dagger M \rangle,
\end{equation}
where $M = \text{diag} (m_u, m_d,m_s)$ and $v = (F^2 m_{\pi^+}^2)/(2m_u + 2m_d)$.
Including the external gauge fields, we have
\begin{align}
D_\mu U &= \partial_\mu U + i U \ell_\mu - i r_\mu U
\\\nonumber
&=\partial_\mu U - i [v_\mu,U] - i \{a_\mu,U\},
\end{align}
with $\ell_\mu \equiv v_\mu - a_\mu$ and $r_\mu \equiv v_\mu + a_\mu$.
The vector and axial-vector couplings appear as
\begin{align}\label{eq:vector_field}
    v_\mu &= G_{V} X_\mu,
    \\
    a_\mu &= G_A X_\mu.
\end{align}
where $G_{V,A} = \, - \text{diag}({g^{V,A}_u,g^{V,A}_d,g^{V,A}_s})$ are the vector and axial-vector coupling matrices.

The Weak interactions lead to $\Delta S = 1$ transitions parameterized in ChPT by the octet and 27-plet operators. 
The former dominates the rates we are interested in and can be written as,
\begin{equation}\label{eq:chpt_deltaS1_app}
    \mathcal{L}_{\Delta S = 1} = F^4 G_8  \langle U_- (D_\mu U)^\dagger D^\mu U \rangle +\text{h.c.},
\end{equation}
where $U_- = (\lambda_6 - i\lambda_7)/2$ is a $3\times 3$ matrix that projects out $s\to d$ transitions. It has one non-zero element, $(U_-)_{32}=1$.
Here, $G_8 =  -\frac{g_8}{\sqrt{2}} G_F V_{us}^* V_{ud}$.
We fix the normalization constant $g_8$ by calculating $K_S \to \pi \pi$, finding $g_8 = 5.1$.
In what follows, we make use of the dimensionless ratio $\kappa_8$ of \cref{eq:kappa_8}.

\cref{eq:chpt_deltaS1_app} induces mixed kaon-pion and kaon-eta propagators.
This multiplies the number of diagrams to be computed for the processes considered here, so we opt to instead diagonalize the kinetic terms~\cite{Ecker:1987hd}. 
In our notation,
\begin{align}
    \pi^+ &\to \pi^+ - \frac{2 \kappa_8}{1-r} K^+,
    \\\nonumber
    K^+ &\to K^+ + \frac{2 \kappa_8 r}{1-r} \pi^+,
    \\\nonumber
    \pi^0 &\to \pi^0 + \frac{2 \kappa_8}{1-r} K_L,
    \\\nonumber
    K_L &\to K_L - \frac{2 \kappa_8 r}{1-r} \pi^0 + \frac{2 \kappa_8 r_\eta}{r_\eta-1} \eta_8,
    \\\nonumber
    \eta_8 &\to \eta_8 - \frac{2 \kappa_8 r_\eta}{r_\eta-1} K_L,
\end{align}
where we also define $r_\eta = m_\eta^2/m_K^2$.
Because we neglect CP violation, no transformation is needed for $K_S$.
As we will see, the eta pole from $K_L-\eta$ mixing does not contribute to the most constraining channels we consider here.
It would, however, contribute to channels such as $K_L\to \gamma \gamma$, $K_L \to e^+e^-$.

\renewcommand{\arraystretch}{1.5}
\begin{table}[t]
    \centering
    \begin{tabular}{l|c||l|c}
        \toprule
        \multicolumn{2}{c||}{$g_i^{V,A} \equiv {\langle I_{0,\pm,3} \,  g^{V,A}\rangle}_{I=T,U,V}$} & \multicolumn{2}{c}{$g_i^{V,A} \equiv \langle{\lambda_i \, g^{V,A}\rangle}$} \\
	\midrule
        $g_{T_0}^{V,A}$ & $g_u^{V,A} + g_d^{V,A}$ & $g_1^{V,A}$ & 0 \\
        $g_{U_0}^{V,A}$ & $g_d^{V,A} + g_s^{V,A}$ & $g_2^{V,A}$ & 0 \\
        $g_{V_0}^{V,A}$ & $g_u^{V,A} + g_s^{V,A}$ & $g_3^{V,A}$ & $g_u^{V,A} - g_d^{V,A}$ 
        \medskip\\
        $g_{T_3}^{V,A}$ & $g_u^{V,A} - g_d^{V,A}$ & $g_4^{V,A}$ & 0 \\
        $g_{U_3}^{V,A}$ & $g_d^{V,A} - g_s^{V,A}$ & $g_5^{V,A}$ & 0 \\
        $g_{V_3}^{V,A}$ & $g_u^{V,A} - g_s^{V,A}$ & $g_6^{V,A}$ & 0 
        \medskip\\
        $g_{T_\pm,U_\pm,V_\pm}^{V,A}$ & 0 & $g_7^{V,A}$ & 0 
        \medskip\\
         &  & $g_8^{V,A}$ & $\frac{1}{\sqrt{3}} \big( g_u^{V,A} + g_d^{V,A} - 2g_s^{V,A}\big) $ \\
        \bottomrule
    \end{tabular}
\caption{A dictionary between couplings in the $T$, $U$, $V$ subspace couplings and the up, down, strange flavor space couplings.
\label{tab:dict}
}
\end{table}

\subsection{Single $X$ vertices}

Expanding the relevant operators in the meson and gauge fields we obtain the relevant interaction vertices for $X$ emission in kaon decay.
When convenient, we collect the following combinations of quark couplings in flavor space,
\begin{equation}
g_i^{V,A} \equiv \langle \lambda_i g^{V,A} \rangle,
\end{equation}
where, for example, $g_3^A = g_u^A - g_d^A$.
Similarly for the $T$, $V$, and $U$-spin flavor subspaces, corresponding to the flavor content of $\pi^+$, $K^+$, and $K^0$, respectively, as shown in \cref{tab:dict}.

In order of increasing number of external fields, the first term we highlight is
\begin{equation}
    \mathcal{L}_{\pi X} =  F X_\mu \left(g_{T_3}^A \partial^\mu \pi^0 + g_8^A \partial^\mu \eta_8  + g_0^A \partial^\mu \eta_0 \right),
\end{equation}
where the last term comes from adding the $\eta_0$ meson to our construction as a singlet state.
The first term above mediates $\pi^0 - X$ transitions through the axial-vector current and appears in the off-shell $X$ contribution to $\pi^0 \to e^+e^-$ decays. 
For on-shell emission of $X$ in meson decays, we can safely neglect such contributions.
They will only show up for off-shell $X_\mu$ suppressed by higher powers of the new-physics couplings.
With the $\Delta S =1$ octet term, an analogous kaon term arises, 
\begin{equation}
    \mathcal{L}_{KX}   =  2 \kappa_8 F \left(2 g_{U_0}^A +\frac{g_{T_3}^A}{1-r} -\frac{1}{\sqrt{3}}\frac{g_{8}^A}{r_\eta-1} \right)  X_\mu(\partial^\mu K_L).
\end{equation}
Note the $g_{T_3}^A$ and $g_8^A$ terms arise due to kinetic mixing with pions and octet eta mesons, respectively.

The $X$ coupling to vector currents is given by
\begin{align}
\label{eq:PiPiX}
    \mathcal{L}_{\pi^2X}   =&  X_\mu \left(g_{T_3}^V \, \pi^+ i\overset{\leftrightarrow}{\partial^\mu} \pi^-\right),
\\
    \mathcal{L}_{K^2X}   =&  X_\mu \left( g_{U_3}^V K_S \overset{\leftrightarrow}{\partial^\mu} K_L + g_{V_3}^V K^+ i\overset{\leftrightarrow}{\partial^\mu} K^- \right) ,
\end{align}
where $A\overset{\leftrightarrow}{\partial_\mu}B = A{\partial_\mu}B - (\partial_\mu A)B$.
In particular, we will also be interested in the $\Delta S= 1$ analogous terms,
\begin{align}\label{eq:XpK}
    \mathcal{L}_{K \pi X}   =& \kappa_8 X_\mu\Bigg[ 
    g_{K\pi^+}^X
    \left(K^+ i\overset{\leftrightarrow}{\partial_\mu}\pi^-  + \text{h.c.}\right) 
     \\\nonumber 
     & \quad -2 g_{K\pi^0}^X(K_S\overset{\leftrightarrow}{\partial_\mu} \pi^0) \Bigg],
\end{align}
where $g_{K\pi^0}^X$ and $g_{K\pi^+}^X$ as defined in \cref{eq:g_kpineutral,eq:g_kpi}.

We now move onto the four-point interactions that contributes to the three-body decays $K \to \pi \pi X$.
First, we will need the $\Delta S =0$ piece, which at $\mathcal{O}(p^2)$ shows up only for kaons or charged pions,
\begin{align}\label{eq:Xppp}
\mathcal{L}_{\pi^3 X}   &= -2 F X_\mu \partial^\mu\pi^0 \Big[g_{T_3}^A \pi^+\pi^-\Big].
\\\label{eq:XKKp}
\mathcal{L}_{K^2 \pi X}   &=  F X_\mu \partial^\mu\pi^0\Bigg[ g_{U_3}^A \frac{(K_L^2 + K_S^2)}{2} 
- g_{V_3}^A (K^+ K^-)\Bigg].
\end{align}
The $\Delta S = 1$ term is given by
\begin{align}\label{eq:XppK}
&\mathcal{L}_{K_L \pi^2 X}   = \frac{\kappa_8 X_\mu}{F} \Bigg[2 g_{K\pi^+\pi^-}^X
(\partial^\mu K_L) (\pi^+\pi^-)
\\\nonumber
&\qquad\qquad - g_{K\pi^0\pi^0}^X(\partial^\mu K_L) (\pi^0\pi^0)\Bigg].
\end{align}
The $K\pi\pi X$ couplings are as defined in \cref{eq:g_kpipiX,eq:g_kpipiX_neutral}.

Despite the fact that the decay $K^+ \to \pi^+ \pi^0$ proceeds only via the $27$-plet operator in the SM, one can still obtain a contribution to the $K^+ \to \pi^+ \pi^0 X_\mu$ through the octet operator.
We find
\begin{align}\label{eq:operator_kpipiX}
   &\mathcal{L}_{K^+ \pi^2 X}   = \frac{\kappa_8 X_\mu}{F} \Bigg[ \left(g_{T_3}^V-\frac{g_{T_3}^A}{3}\right)  (\partial^\mu K^+) \pi^-\pi^0
\\\nonumber
&- g_{K\pi^+\pi^0}^X  K^+ (\pi^-\overset{\leftrightarrow}{\partial}_\mu\pi^0) + \text{h.c.}\Bigg],
\end{align}
where the coupling $g_{K\pi^+\pi^0}^X$ is defined \cref{eq:Kplus_piplus_pi0_X}.
Note that in this case there are two operators at play in the $K^+ \to \pi^+ \pi^0 X$ decay, namely a $\pi^+ \leftrightarrow \pi^-$ symmetric and anti-symmetric piece, which correspond to two independent amplitudes.

\subsection{Double $X$ vertices}

Now, we consider terms with two insertions of $X_\mu$.
First, the $\Delta S = 0$ terms,
\begin{align}
    \mathcal{L}_{\pi^2 X^2} &= 4 X_\mu X^\mu \Bigg[
    \left(g_{T_3}^{L}g_{T_3}^{R}\right)(\pi^+\pi^-)
    \Bigg],
\end{align}
where the $X_\mu X^\mu \pi^0\pi^0$ is absent at $\mathcal{O}(p^2)$ but shows up at $\mathcal{O}(p^4)$. 
For the kaons,
\begin{align}
    \mathcal{L}_{K^2 X^2} &= 2 X_\mu X^\mu \Bigg[
    \left(g_{U_3}^{L}g_{U_3}^{R}\right)(K_L^2 + K_S^2)
    \\\nonumber 
    & \quad +2\left(g_{V_3}^{L}g_{V_3}^{R}\right)(K^+K^-)
    \Bigg],
\end{align}
where we note the $L\times R$ structure of the coupling.
Now the $\Delta S = 1$ terms,
\begin{align}
    \mathcal{L}_{K\pi X^2} &=  2 \kappa_8 X_\mu X^\mu \Bigg[ g_{K\pi^+}^{XX}(K^+\pi^-+\text{h.c.})   - 2g_{K\pi^0}^{XX} K_L \pi^0
    \Bigg],
\end{align}
where as expected, all vertices vanish for a conserved vector current. 

\subsection{Mixed $X\gamma$ vertices}

To obtain purely EM vertices, one can simply make use of the $X_\mu$ interactions above with $g^A \to 0$ and $g^V_{u,d,s} = Q_{u,d,s}$, with $Q$ the respective quark electric charge.
For mixed interactions, we include the photon and the dark vector simultaneously in ChPT by adding the photon to the vector field in \cref{eq:vector_field},
\begin{align}
    v_\mu &\to v_\mu - e A_\mu \left(\frac{\lambda_3}{2} +  \frac{\lambda_{8}}{2\sqrt{3}}\right) 
    \\\nonumber &= v_\mu -\frac{eA_\mu}{3} \text{diag}(2,-1,-1).
\end{align}

To calculate $K\to \pi \gamma X$, we will need
\begin{align}
    \mathcal{L}_{K\pi X\gamma} &= 2 \kappa_8 e X_\mu A^\mu g_{K\pi^+}^X (K^+ \pi^- + \text{h.c.}),
\end{align}
as well as
\begin{align}
    \mathcal{L}_{(K^2,\pi^2) X\gamma} &= 2 e X_\mu A^\mu (g_{T_3}^V\pi^+ \pi^- 
    + g_{V_3}^V K^+ K^-).
\end{align}
Note that as per \cref{eq:XpK}, the $K\pi \gamma$ vertex vanishes.

\bibliographystyle{utphys}
\bibliography{ref}

\end{document}